\documentclass[12pt]{article}

\usepackage{graphicx}
\begin{document}

\begin{center}
{\large \bf Super strong nuclear force caused by migrating $\bar{K}$ mesons\\
- Revival of the Heitler-London-Heisenberg scheme in kaonic nuclear clusters
}

\vspace{1cm}
(Published in Proc. Jpn. Acad. {\bf B 83} (2007) 144;\\
http://www.jstage.jst.go.jp/browse/pjab)

\vspace*{0.5cm}
Toshimitsu Yamazaki$^{a)}$ and Yoshinori Akaishi$^{b)}$\\
 \vspace*{2mm}
 {\it a) Department of Physics, University of Tokyo, Bunkyo-ku, Tokyo 113-0033, Japan, and
RIKEN Nishina Center, Wako, Saitama 351-0198, Japan,\\

 b) College of Science and Technology, Nihon University, Funabashi, Chiba 274-8501, Japan, and RIKEN Nishina Center, Wako, Saitama 351-0198, Japan}
\end{center}

\begin{abstract}
 We have studied the structure of $K^- pp$ comprehensively by solving this three-body system in a variational method, starting from the Ansatz that the $\Lambda(1405)$ resonance $(\equiv \Lambda^*)$ is a $K^-p$ bound state. The structure of $K^-pp$ reveals a molecular feature, namely, the $K^-$ in $\Lambda^*$ as an ``atomic center" plays a key role in producing strong covalent bonding with the other proton. We point out that strongly bound $\bar{K}$ nuclear systems are formed by ``super strong" nuclear force due to migrating real bosonic particles $\bar{K}$ {\it a la} Heitler-London-Heisenberg, whereas the normal nuclear force is caused by mediating virtual pions. We have shown that the elementary process, $p + p \rightarrow K^+ + \Lambda^* + p$, 
which occurs in a short impact parameter and with a large momentum transfer, leads to unusually large self-trapping of $\Lambda^*$ by the involved proton, since the $\Lambda^*$-$p$ system exists as a compact doorway state propagating to $K^-pp$.
\end{abstract}

\newcommand{\kbar}{$\bar{K}$}
\newcommand{\km}{$K^-$}


\section{Introduction}

In 1932, right after the discovery of the neutron, Heisenberg \cite{Heisenberg:32} tried to explain the nuclear force (for instance, the proton-neutron interaction) with the idea of ``Platzwechsel" of a migrating particle, which had been known as the mechanism for the covalency in hydrogen molecule, first clarified by Heitler and London in 1927 \cite{Heitler:27}. This ``molecule-type bonding"  mechanism can be written as 
\begin{equation}
{\it Heitler-London-Heisenberg:}~~e^- p + p \leftrightarrow p + e^- p. \label{eq:1-1}
\end{equation}
Since the ``$e^- p$" cannot be identified to the neutron for obvious reasons, this  idea was unsuccessful, and was abandoned. Instead of a ``migrating real" particle, Yukawa \cite{Yukawa:35} introduced a ``mediating virtual" boson for explaining the nuclear force as
\begin{equation}
{\it Nuclear~force~by~ Yukawa:}~~p \leftrightarrow \pi^+ + n,~~n \leftrightarrow \pi^- + p. \label{eq:1-2}
\end{equation}
This hypothetical particle was eventually discovered, and Yukawa' s idea of a mediating virtual particle   was established as the fundamental concept for understanding all the forces including the electroweak interaction. In the present paper we point out that the $\bar{K}$ meson, as a ``real migrating" particle,  plays a unique role in producing strong bonding of nucleons as
\begin{equation}
{\it Super ~strong~nuclear~ force:}~~K^- p + p \leftrightarrow p + K^- p. \label{eq:1-3}
\end{equation}
One can say that this is the revival of the forgotten Heitler-London-Heisenberg scheme for nuclear  binding force. In the following sections we describe briefly how this view has come out from our theoretical studies of an exotic bound system $K^-pp$. We also foresee perspectives of this view toward kaon condensation. Full accounts of the results will be published soon \cite{Yamazaki:07a,Yamazaki:07b}.   

\section{Kaonic nuclear bound states}

Recently, exotic light nuclear systems involving a $\bar{K}$  ($K^-$ and $\bar{K}^0$) meson as a constituent have been predicted based on phenomenologically constructed $\bar{K}N$ interactions \cite{Akaishi:02,Yamazaki:02,Dote:04,Yamazaki:04,Akaishi:05,Kienle:06}. The basic ingredient for this new family of nuclear states, often called {\it kaonic nuclear clusters}, is the strongly attractive $I=0$ $\bar{K} N$ interaction, which accommodates the $K^- p$ bound state, identified to the known $\Lambda (1405)$ resonance in the $\Sigma\pi$ channel (hereafter, expressed as $\Lambda^*$) with a binding energy of $B_K$ = 27 MeV and a width of $\Gamma$ = 40 MeV \cite{Dalitz}. Since the $\Lambda (1405)$ resonance is largely populated in the $K^-$ absorption at rest in $^4$He ~\cite{Riley:75} and also in nuclear emulsion \cite{Davis:77}, it is very likely to be the $I=0~\bar{K} N$ bound state. The $\bar{K} N$ interaction was also derived theoretically (see, for instance, Refs.  \cite{Mueller:90,Weise:96}).

The most spectacular property of the predicted kaonic nuclei is in their extremely high densities; the average density, $\rho_{av} \sim 0.5$ fm$^{-3}$, reaches about 3 times as much as the normal nuclear density $\rho_0  \sim 0.17$ fm$^{-3}$. This is an enormous contrast to the normal nuclear systems, in which the nuclear density is always a constant. The $\bar{K}$ produces extra binding of nucleons, which overcompensate the stiff nuclear incompressibility. In order to understand this feature we study the most fundamental unit, $K^-pp$.

The lightest system following the ``$\Lambda(1405) = K^-p$ Ansatz", $K^-pp$, was predicted to exist with $M = 2322$ MeV/$c^2$, $B_K$ = 48 MeV and $\Gamma$ = 61 MeV  \cite{Yamazaki:02}. Recently, Faddeev calculations have been carried out to obtain the pole corresponding to $K^-pp$ by Shevchenko {\it et al.} \cite{Shevchenko:06} and by Ikeda and Sato \cite{Ikeda:06}. Their results are consistent with ours. 
\section{Structure of $K^- pp$}\label{sec:structure}

In our study the ATMS variational method \cite{Akaishi:86} was employed together with the bare $\bar{K} N$ interaction of AY \cite{Akaishi:02} and the bare $NN$ interaction of Tamagaki \cite{Tamagaki}.
The three-body variational wave function of $\bar{K}NN$ with a number definition $(1, 2, 3) = (\bar{K}, N, N)$ is given as 
\begin{equation}
\Psi = [\Phi_{12} + \Phi_{13}]\,|T=1/2>
\end{equation}
where
\begin{eqnarray} \label{eq:Phi}
\Phi_{12} &=& [f^{I=0} (r_{12})\, P_{12}^{I=0}  + f^{I=1} (r_{12})\, P_{12}^{I=1}]
\times f_{NN} (r_{23}) f(r_{31}),\\
\Phi_{13} &=& f (r_{12}) f_{NN}(r_{23})
\times [f^{I=0} (r_{31})\, P_{31}^{I=0}  + f^{I=1} (r_{31})\, P_{31}^{I=1}], 
\end{eqnarray}
with
$P_{12}^{I=0} = (1 - \vec{\tau_K} \cdot \vec{\tau_N})/4$ and
 $P_{12}^{I=1} = (3 + \vec{\tau_K} \cdot \vec{\tau_N})/4$.
The functions $f^{I=0} (r_{ij})$ and $f^{I=1} (r_{ij})$ are scattering correlation functions of the particle pair $(i,j)$ for the $I=0$ and $I=1$ $\bar{K}N$ interactions, respectively, and $f_{NN} (r_{23})$ is that for the $NN$ pair, and $f(r_{i,j})$ is for the off-shell case.  
The $T=1/2$ state consists of two isospin eigenstates as
\begin{equation}
|T=1/2> =  \sqrt{\frac{3}{4}} \, \biggl[(\bar{K}_1 N_2)^{0,0} \, p_3 \biggr] + \sqrt{\frac{1}{4}} \, \biggl[ - \sqrt{\frac{1}{3}} (\bar{K}_1 N_2)^{1,0} \, p_3 
+ \sqrt{\frac{2}{3}} (\bar{K}_1 N_2)^{1,1} \, n_3 \biggr] , 
\end{equation}
where $(\bar{K}_1 N_2)^{I, I_z}$ is for the isospin $(I, I_z)$. Among these the first term corresponds to $\Lambda^* p$.

\begin{figure}[htb]
\centering
\includegraphics[width=14cm]{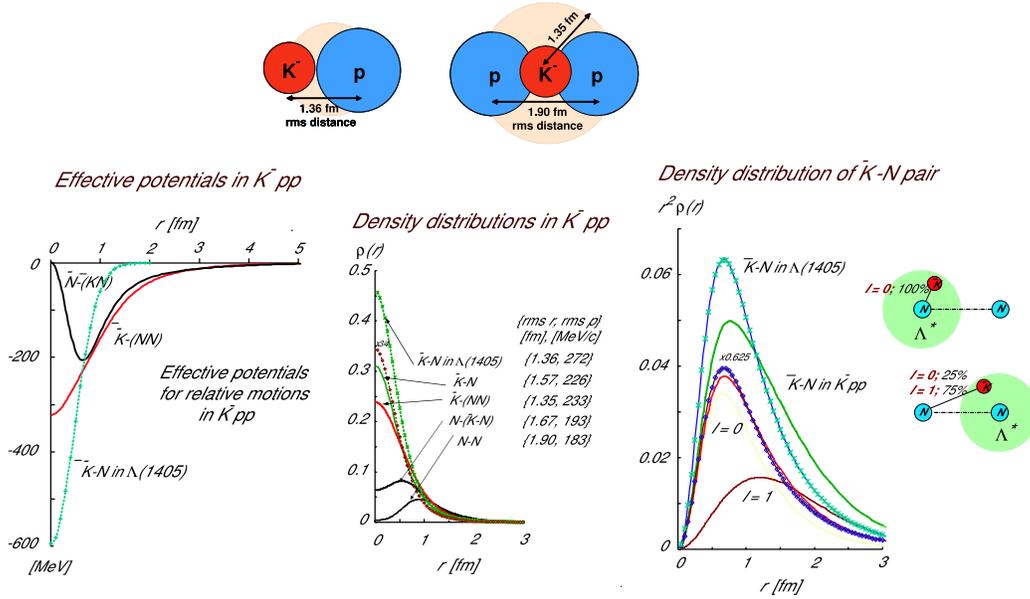}
\vspace{0cm}
\caption{\label{fig:Structure-Kpp} 
(Upper) Schematic structure of $K^-p$ and $K^-pp$. 
(Lower Left) The effective potentials for relative motions of $N$-$(\bar{K}N)$ and $\bar{K}$-$(NN)$, deduced from the exact variational wavefunction for $K^-pp$. The $\bar{K}$-$N$ potential in $\Lambda(1405)$ is also shown. (Lower Middle) Density distributions of various coordinates in $K^-pp$ as well as $\Lambda(1405) = K^-p$. (Lower Right) Comparison of the density distributions, $r^2 \rho(r_{KN})$, of the $\bar{K}$-$N$ distance in the  $\bar{K} N$ pair in $\Lambda (1405)$ and in $K^-pp$. The latter is decomposed into the $I=0$ and $I=1$ pairs. The density distribution in $\Lambda(1405)$ after multiplication of a factor 0.625 is also shown. }
\end{figure}

The predicted structure of $K^-p$ and $K^-pp$ is shown in Fig.~\ref{fig:Structure-Kpp}. 
The ``nucleus" $pp$ does not exist, but the $K^-$ can combine two protons into a strongly bound system, when they are in a spin-singlet state.  

The effective potential energies as functions of the relative distances of $\bar{K}$-$(NN)$ and $N$-$(\bar{K} N)$ are extracted from the obtained total wave function, as shown in Fig.~\ref{fig:Structure-Kpp} (Left). The distributions of the relative distances and the momenta of the constituent particle pairs, namely, 
$\bar{K}$-$N$, $\bar{K}$-$(NN)$, $(\bar{K} N)$-$N$, and $N$-$N$, were calculated, as shown in the middle. For comparison the $\bar{K}$-$N$ density distribution in free $\Lambda^*$ is also shown. The $N$-$N$ rms distance is 1.90 fm, which is significantly smaller than the average inter-nucleon distance in normal nuclei (2.2 fm \cite{Bethe:71}), and is much smaller than the rms distance of $p$-$n$ in $d$ (3.90 fm). The rms radius of $\bar{K}$ with respect to $(NN)$ is 1.35 fm, close to the rms distance of $\bar{K}$-$N$ in $\Lambda(1405)$, 1.36 fm.

We compare in Fig.~\ref{fig:Structure-Kpp} (Right) the $\bar{K}$-$N$ distance distribution of the $\bar{K} N$ pair in $K^-pp$, $\rho_{\bar{K}-N}(K^-pp)$, with that in $\Lambda(1405)$, $\rho_{\bar{K}-N}(\Lambda^*)$. The former ($R^{\rm rms}_ {\bar{K}-N}$ = 1.57 fm) is significantly broader than the latter (1.36 fm). We decompose the density distribution into the $\bar{K} N^{I=0}$ and $\bar{K} N^{I=1}$ parts, as shown. The $I=0$-pair distribution has a shape closer to $\rho_{\bar{K}-N}(\Lambda^*)$, whereas the $I=1$ part is widely distributed due to the smaller attractive interaction.  
When $K^-$ (1) resides with Proton (2) with a probability of 0.5, the $I=0$ component of the wave function $\Phi_{12}$ in (\ref{eq:Phi}) dynamically increases to 1 due to the strong $\bar{K} N^{I=0}$ interaction. Adding an intensity ($0.5 \times 1/4 = 0.125$) from Proton (3), we expect the total intensity to be $0.625 \, \rho_{\bar{K}-N}(\Lambda^*)$, which accounts for the calculated $\rho_{\bar{K}-N}(K^-pp)$ very well. This means that $K^-$ (1) in $K^-pp$ resides partially around Proton (2) in a form of $\Lambda (1405)$, and partially around Proton (3), as given by the total wave function. This indicates that the structure of $\Lambda(1405)$ is nearly unchanged when it dissolves into this ``nucleus". In other words, the $\Lambda(1405)$ state, though modified, persists in a nuclear system. This aspect  justifies the $\Lambda(1405)$ doorway model \cite{Yamazaki:02}.

\begin{figure}[htb]
\centering
\includegraphics[width=12cm]{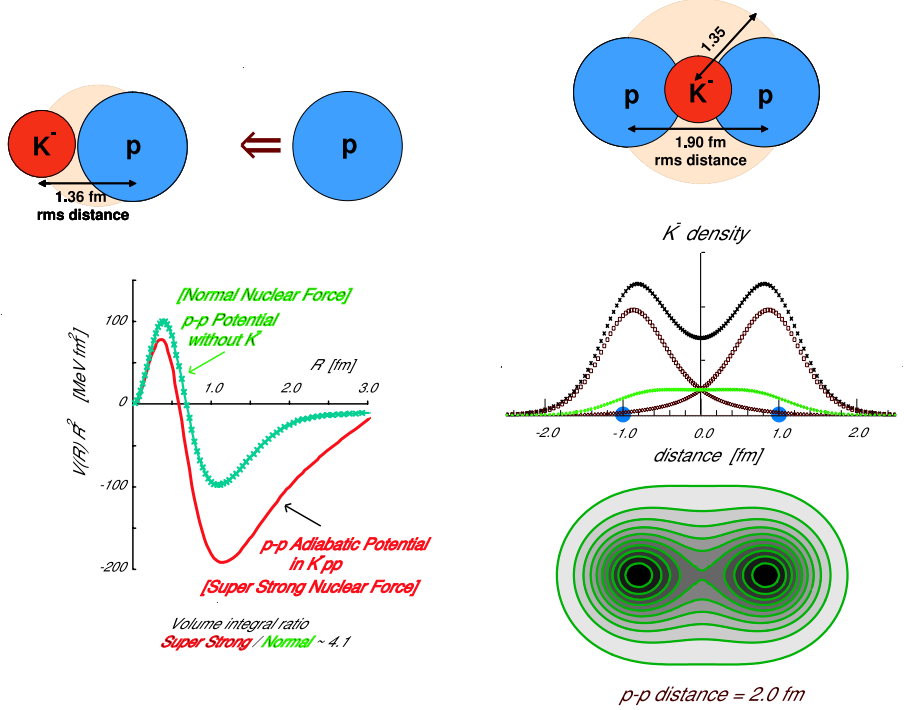}
\vspace{0cm}
\caption{\label{fig:AdiaPot} 
(Left) The adiabatic potential ($V(R) \, R^2$), when a proton approaches a bound $K^-p$ ``atom" ($\Lambda^*$), as a function of the distance between $p$ and $p$. For comparison the Tamagaki potential for the normal $V_{NN}$ interaction is shown.
(Right) The molecular structure of $K^-pp$. The projected density distributions of $K^-$ in $K^-pp$ with a fixed $p-p$ distance (= 2.0 fm) and the corresponding $K^-$ contour distribution are shown, }
\end{figure}

\section{Molecular aspect of $K^-pp$ and super strong nuclear force}

The present kaonic nuclear cluster $K^-pp$ can be interpreted as a kaonic hydrogen molecular ion in the sense that $K^-$ migrates between the two protons, producing ``strong covalency" through the strongly attractive $\bar{K}N^{I=0}$ interaction. This is essentially the mechanism of Heitler and London \cite{Heitler:27} for hydrogen molecule, though the nature of the interaction is totally different and the migrating particle is much heavier and bosonic.  Figure~\ref{fig:AdiaPot} (Left) shows the adiabatic potential ($V(R) \, R^2$), when a proton approaches a $\Lambda(1405)$ particle, as a function of the $p$-$p$ distance. The deep potential indicates that a proton approaching from a distant place to an isolated $\Lambda^*$ gets quickly trapped and dissolved into the bound state of $K^-pp$. 

Figure.~\ref{fig:AdiaPot} (Right) shows the projected distribution of $K^-$ along the $p$-$p$ axis and the contour distribition of $K^-$, when the $p$-$p$ distance is fixed to 2.0 fm (this case resembles the ground state of $K^-pp$, as the calculated rms distance is 1.9 fm). The $K^-$ is distributed, not around the center of $p$-$p$, but around each of the two protons. The $K^-$ distribution is composed of the ``atomic" part, as shown by brown dotted curves, and the exchange part (green broken curve) {\it a la} Heitler and London. 

We emphasize that the strong $I=0$ $\bar{K}N$ attraction produces a large exchange integral,
\begin{equation}
\sum _{\{i,j\} = \{2,3\},\{3,2\}} <\Phi_{1i} |v_{\bar{K}N} (12) + v_{\bar{K}N} (13) |\Phi_{1j}>  = -52.6~{\rm MeV},
\end{equation}
which is the source for the deeper binding of $K^-pp$ as compared with $K^-p$.
Despite the drastic dynamical change of the system caused by the strong $\bar{K} N$ interaction the identity of the ``constituent atom", $\Lambda^*$, is nearly preserved because of the presence of a short-range repulsion between the two protons. The molecule $K^-pp$ resembles a tightly bound $\Lambda^*$-$p$, which we call $\Lambda^* p$ doorway in the formation process.  

Thus, we have demonstrated that the strong $\bar{K} N$ attraction produces a very strong molecular type bonding of the two protons. This adiabatic potential (red curve) as shown in Fig~\ref{fig:AdiaPot} (Left) can be called {\it Super Strong Nuclear Force}, as compared with the ordinary nuclear force (green curve). Not only the depth of the potential, but also the long range attractive part due to the covalent nature and the relatively smaller short-range repulsive part produce an enormous binding in $\bar{K}$-migrating nuclear systems, as shown in Fig~\ref{fig:AdiaPot} (Left). The ``super strong" / ``normal" ratio of the volume integrated strength is 4.1.

\section{Dominance of $\Lambda^* p$ doorway in $K^-pp$ production in $NN$ collisions}\label{sec:pp-reaction}

\begin{figure}[h]
\centering
\includegraphics[width=14cm]{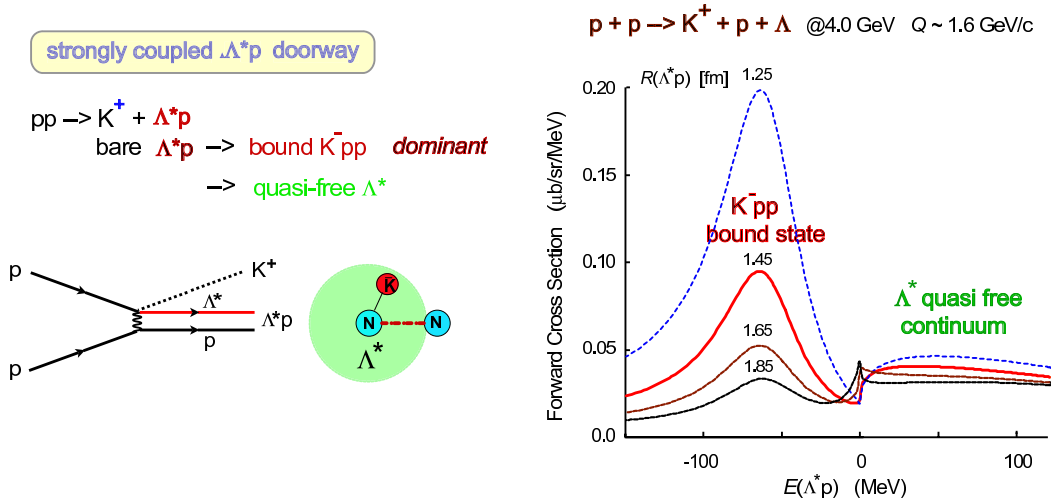}
\vspace{0cm}
\caption{\label{fig:pp2KX} 
 (Left) Diagram for the $p(p,K^+)K^-pp$ reaction. (Right) Calculated spectral shape for different rms distances $R(\Lambda^*$-$p)$, arbitrarily chosen. The binding energy of $K^-pp$ is set to be 86 MeV for the $\bar{K} N$ interaction which is 17 \% enhanced.  
 In this case, $R(\Lambda^* p) = 1.45$ fm is realistic. }
\end{figure}

We expect that the $K^-pp$ state as a $\Lambda^* p$ composite can be formed  in the $p+p$ reaction: 
\begin{equation}
p + p  \rightarrow K^+ + (\Lambda^* p)
         \rightarrow  K^+ + K^-pp.\label{eq:pp2KLp}
\end{equation}
The reaction diagram is shown in Fig.~\ref{fig:pp2KX} (Left). 
 Essentially, the spectral function for $= K^- pp$ is composed of the following three factors:
 i) the collision range $1/m_B$, taken to be the $\rho$ meson mass; $m_B =  m_{\rho} =  770$ MeV/$c^2$. ii) the momentum transfer, $Q \sim 1.6$ GeV/c, and iii) the structure function, depending on the rms distance $R(\Lambda^* p)$ of the $\Lambda^*$-$p$ system. 
 
The calculated spectral function at $T_p$ = 4 GeV at forward angle in the scale of $E(\Lambda^* p) = 27~{\rm MeV} - B_K$ is presented in Fig.~\ref{fig:pp2KX} (Right). Surprisingly, in great contrast to the ordinary reactions, the spectral function is peaked at the bound state with only a small quasi-free component. This means that the sticking of $\Lambda^*$ and $p$ is extraordinarily large.
This dominance of $\Lambda^* p$ sticking in such a large-$Q$ reaction can be understood as originating from the matching of the small impact parameter with the small size of the bound state. 
It is vitally important to examine our results experimentally. An experimental observation of $K^- pp$ in $pp$ collision may be revealed in a past experiment DISTO at Saclay \cite{DISTO} as well as in a planned experiment at GSI \cite{FOPI}. If a bound-state peak is found, it will not only confirm the existence of $K^- pp$, but also prove the compactness of the $\bar{K}$ cluster.

\section{Why does $\bar{K}$ produce shrunk nuclear systems?}

The constancy of nuclear density, $\rho \sim \rho_0 \sim 0.17$ fm$^{-3}$, which is believed to be almost the nuclear physics ``law", results from the balance between the short-range nuclear repulsion and the long-range attraction of the ordinary nuclear force. 
In normal nuclei the average inter-nucleon distance is 
\begin{equation}
d_{NN} \sim 2 \, \sqrt[3]{\frac{3}{4 \pi \rho_0}} \sim 2.2~{\rm fm},
\end{equation}
when we adopt Bethe's estimate based on a close-packed sphere approximation \cite{Bethe:71}. The nucleon rms radius, $r_{\rm rms} \sim 0.86$ fm, corresponds to a nucleon volume of $v_N \sim$ 2.66 fm$^3$ and to a nucleon density of $\rho_N \sim$ 0.38 fm$^{-3}$. This means that nucleons occupy the nuclear space with a compaction factor of $f_c = \rho_N/\rho_0 \sim$ 2.3. This situation seems to be hardly changed by the normal nuclear force.

The hard core part of the $N$-$N$ interaction plays an essential role in keeping the nuclear density constant. In an intuitive picture it is related to the Pauli blocking in the $u-d$ quark sector. This situation is common for all hadron-hadron interactions except for the case of $\bar{K}$, which, composed of $s\bar{u}$ ($\bar{K}^0$) or $s\bar{d}$ ($K^-$), includes no $(u,d)$ quark. Thus, the $\bar{K} N$ interaction is expected to be dominated by the $u-\bar{u}$ and $d-\bar{d}$ attraction without short-range repulsion. From this consideration we understand intuitively why normal nuclei are difficult to compress and why only the $\bar{K}$ meson produces dense nuclear systems. The intruding $\bar{K}$ meson, as a ``messenger of $\bar{u}/\bar{d}$ quark", is expected not only to cause super strong nuclear force by its strong binding and migration with protons, but also to relax the $NN$ repulsion due to a kind of  ``anti-quark shielding" effect: $uud -s\bar{u} - uud$. The latter effect is not taken into account in our calculation, where the nucleons are  treated as structureless elementary particles, but will certainly enhance the super strong nuclear force. 
 
When the nuclear density exceeds the nucleon compaction factor, an additional effect may come in, because the QCD vacuum is expected to vanish and chiral symmetry is restored \cite{Nambu,Hatsuda:85,Vogl:91}. It is vitally important to investigate to what extent the involved hadrons keep their identities under such an extremely dense system. 

\begin{figure}[h]
\centering
\includegraphics[width=6cm]{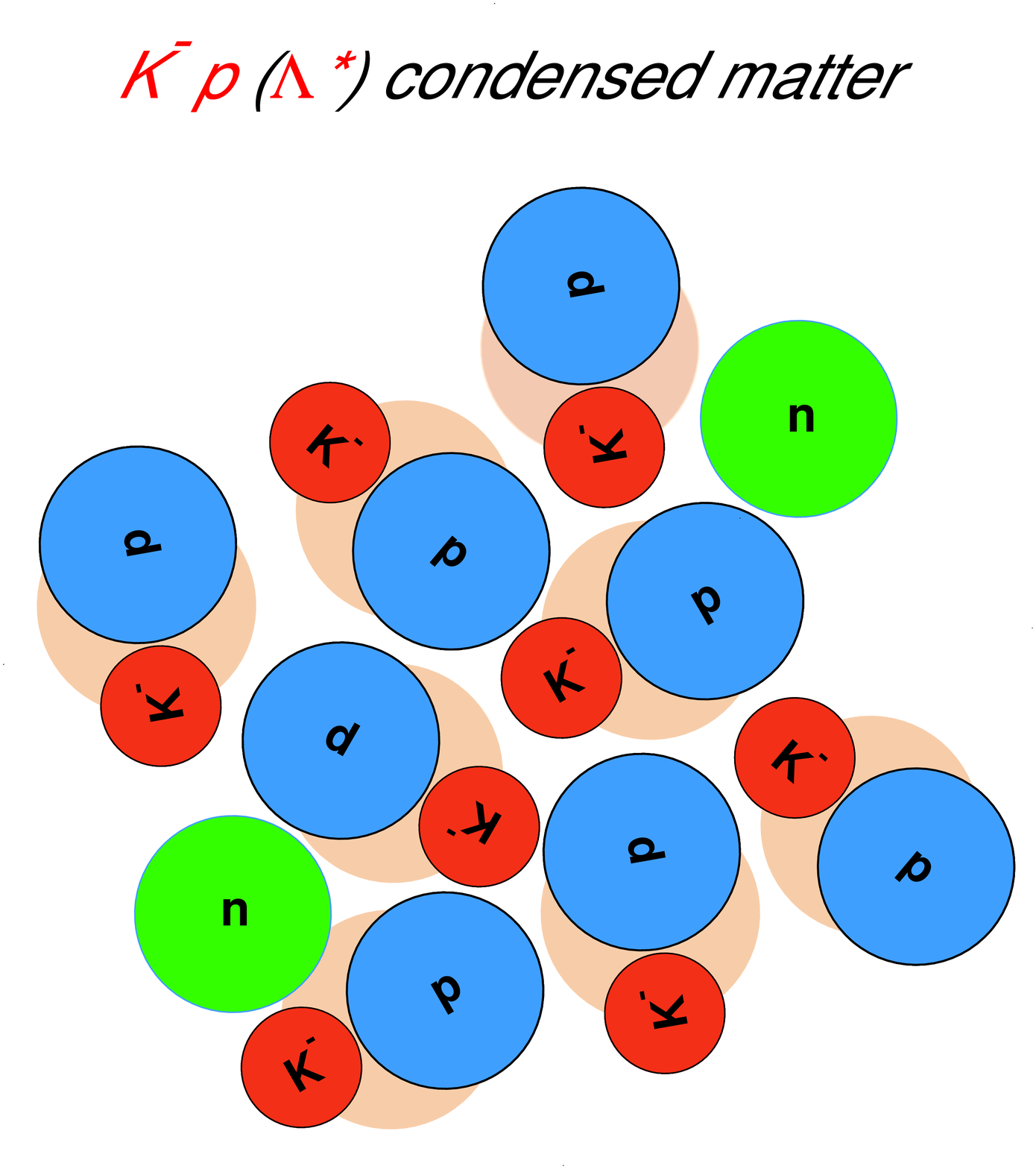}
\includegraphics[width=6cm]{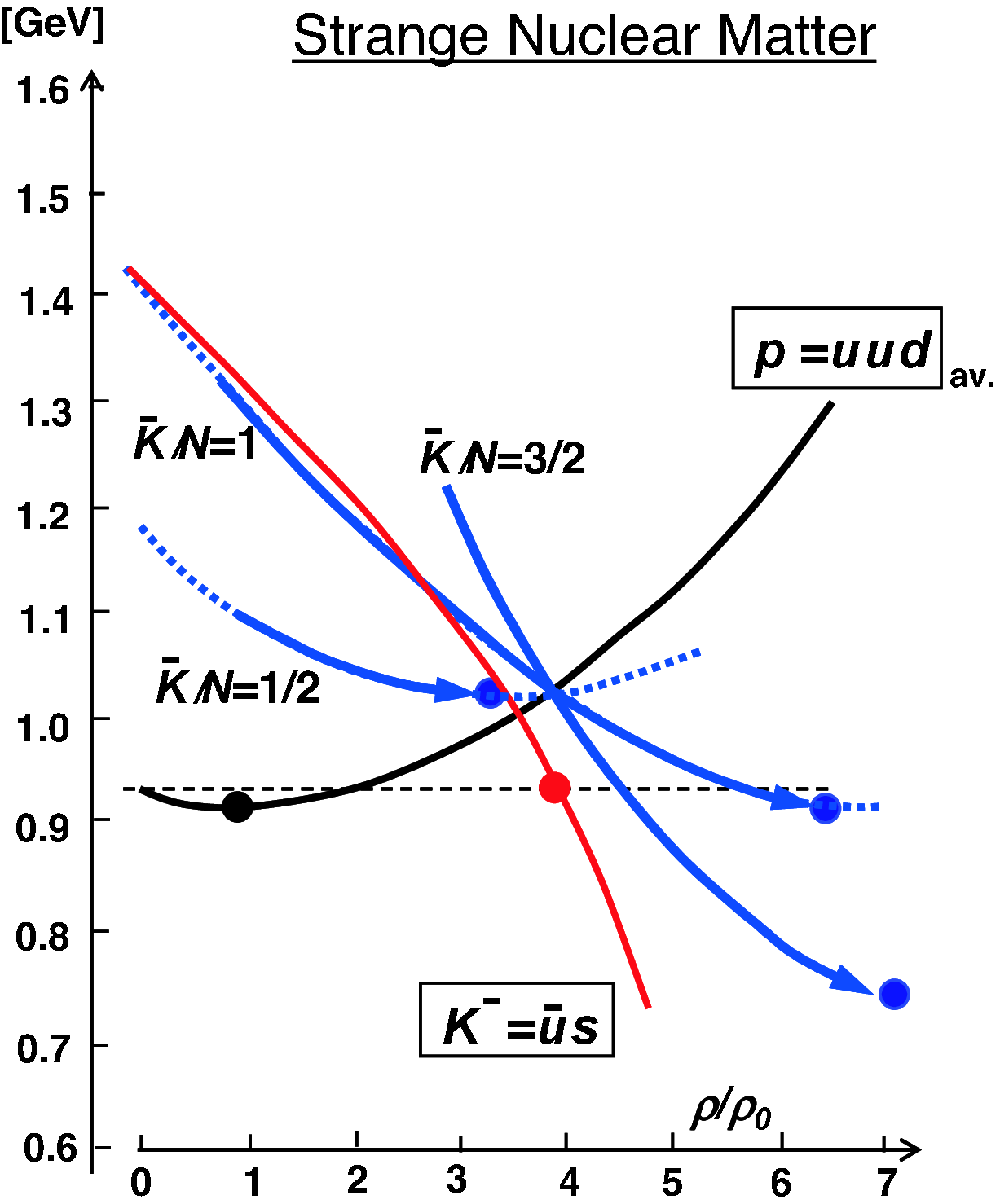}
\vspace{0cm}
\caption{\label{fig:Kmatter} (Left) Speculated $K^-p = \Lambda^*$ matter with a quasi-$\Lambda^*$  as an ``atomic" constituent, where $K^-$'s migrate among protons, producing high-density strange matter. (Right) Speculated diagrams for the density dependences of the bound-state energies of various baryon composite systems ($pK^-)^{m}$n$^{n}$. The $\bar{K} N$ energy is represented by the red curve, and the nuclear compression by the black curve. The total energies for representative fractions of $K^-/N$ (=1/2, 1 and 3/2) are depicted by respective blue curves, showing minima at high density and low energy. Density-dependent enhanced $\bar{K}N$ interactions with relativistic correction are assumed.
}
\end{figure}

\section{The super strong nuclear force toward kaon condensation}

We have studied many other kaonic bound states  \cite{Dote:04}. The strong bonding produced by a single $\bar{K}$ was shown to saturate for 3-4 nucleons.  
We have also predicted the double-$K^-$ clusters \cite{Yamazaki:04}. The species, $K^-K^-pp$, corresponds to the neutral hydrogen molecule, and has a large binding energy $B_{KK} = 117$ MeV and a shrunk inter-nucleon distance, $R_{N-N}^{\rm rms} = 1.3$ fm \cite{Yamazaki:04}. Expecting that the kaonic bound states with a large number of $\bar{K}$ will be more deeply bound and stable, we proposed to search for them in high-energy heavy reactions \cite{Yamazaki:04}. 

We can conceive a multi-$\bar{K}$ system, as sketched in Fig.~\ref{fig:Kmatter} (Left), where $K^-$ mesons migrate coherently among protons. We speculate that the ground state of such a system shows a large energy gap. The above consideration naturally leads us to a regime of kaon condensation \cite{Kaplan:86,Brown:94}. Namely, $\bar{K}$ mesons, as intruders with $\bar{u}$ and $\bar{d}$ quarks, behave as strong glue to combine surrounding nucleons to a dense system. The whole energy drops down, depending on the composition of $p$, $n$ and $\bar{K}$. Intuitively, one can conceive energy diagrams as shown in Fig.~\ref{fig:Kmatter} (Right). There are three phases: the $\bar{K}$ matter with energy $E_K (\rho)$, the $\Lambda$  matter with energy $E_{\Lambda} (\rho)$ and the normal nuclear matter with $E_N (\rho)$. Only the $\bar{K}$  matter decreases in energy as the density increases. Then, there are three regimes for the phase stability.
\begin{eqnarray}
 i)~ &&E_K (\rho) > E_{\Lambda} (\rho):~~{\rm strong~ decay~to~}\Lambda'{\rm s} \\
ii)~&& E_N (\rho) < E_K (\rho) < E_{\Lambda} (\rho):~{\rm  weak~ decay:}~\tau \sim {\rm sub-ns} \\
iii)~&& E_N (\rho) >  E_K (\rho):~~{\rm stable}
\end{eqnarray}
Kaonic nuclear clusters in the cases i) and ii) can be searched for in laboratory. The case iii) corresponds to a stable kaonic matter which might exist as a strange star.\\

\begin{figure}[h]
\centering
\includegraphics[width=9cm]{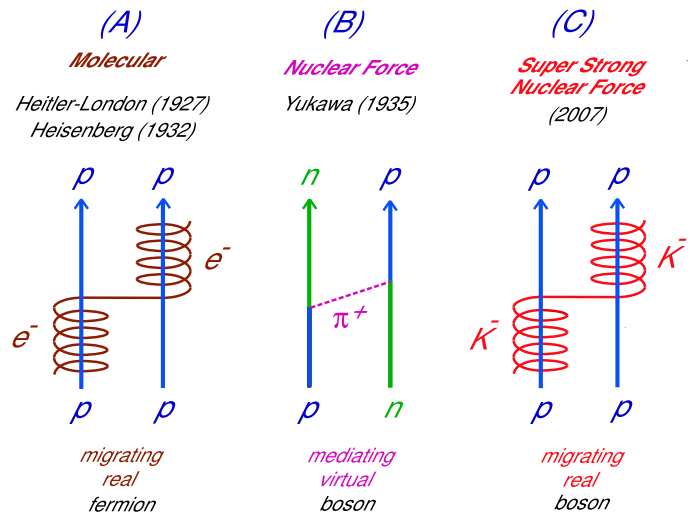}
\vspace{0cm}
\caption{\label{fig:SuperStrong} Summary of the three different interaction schemes for nuclear forces. (A) The Heitler-LOndon-Heisenberg model. (B) The Yukawa interaction. (C) The super strong nuclear force by the $\bar{K}$ covalency. }
\end{figure}

In the present paper we have clarified that the $\bar{K}$ meson can produce an enormously strong nucleon-nucleon binding by the Heitler-London-Heisenberg mechanism. The $\bar{K}$-migrating $NN$ bonding is deeper and longer-ranged  with relatively smaller short-range repulsion. This super strong nuclear force can make ultra dense nuclear systems without the aid of gravity. Figure \ref{fig:SuperStrong} summarizes the three kinds od nuclear forces.

\section*{Acknowledgements}
We would like to thank Prof. P. Kienle for the stimulating discussion. One of us (T.Y.) is grateful to the Alexander von Humboldt Foundation for its ``Forschungspreis".  We acknowledge the receipt of Grant-in-Aid for Scientific Research of Monbu-Kagakusho of Japan.


\begin{thebibliography}{99}
\bibitem{Heisenberg:32} W. Heisenberg, Z. Phys. {\bf 77} (1932) 1, {\bf 78} (1932) 156, {\bf 80} (1933) 587.
\bibitem{Heitler:27} W. Heitler and F. London, Z. Phys. {\bf 44} (1927) 455.
\bibitem{Yukawa:35}
H. Yukawa, Proc. Phys. Math. Soc. Jap. {\bf 17} (1935) 48.
\bibitem{Yamazaki:07a} T. Yamazaki and Y. Akaishi, arXiv/nucl-th/0604049; Proc. YKIS Workshop (Kyoto, 2006); Prog. Theor. Phys., in press.
\bibitem{Yamazaki:07b} T. Yamazaki and Y. Akaishi, submitted to Phys. Rev. C.
\bibitem{Akaishi:02} Y. Akaishi and T. Yamazaki, Phys. Rev. {\bf C 65} (2002) 044005.
\bibitem{Yamazaki:02} T. Yamazaki and Y. Akaishi, Phys. Lett. {\bf B 535} (2002) 70.
\bibitem{Dote:04} A.~Dot$\acute{\rm e}$, H. Horiuchi, Y. Akaishi and T. Yamazaki, Phys. Lett. {\bf B 590} (2004) 51;
A.~Dot$\acute{\rm e}$, H. Horiuchi, Y. Akaishi and T. Yamazaki, Phys. Rev. {\bf C 70} (2004) 044313.
\bibitem{Yamazaki:04} T. Yamazaki, A.~Dot$\acute{\rm e}$ and Y. Akaishi, Phys. Lett. {\bf B 587} (2004) 167. 
\bibitem{Akaishi:05} Y. Akaishi, A.~Dot$\acute{\rm e}$ and T. Yamazaki, Phys. Lett. {\bf B 613} (2005) 140.
\bibitem{Kienle:06} P. Kienle, Y. Akaishi and T. Yamazaki, Phys. Lett. {\bf B 632} (2006) 187.
\bibitem{Dalitz} R.H. Dalitz and S.F. Tuan, Ann. Phys. {\bf 8} (1959) 100; {\bf 10} (1960); Phys. Rev. Lett. {\bf 2} (1959) 425; R.H. Dalitz, Rev. Mod. Phys. {\bf 33} (1961) 471; Phys. Rev. Lett. {\bf 6} (1961) 239. 
\bibitem{Riley:75} B. Riley, I-T. Wang, J.G. Fetkovich and J.M. McKenzie, Phys. Rev. {\bf D11} (1975) 3065.
\bibitem{Davis:77} D.H. Davis, D.N. Tovee and R. Nowak, Nukleonika {\bf 22} (1977) 845.
\bibitem{Mueller:90} A. M\"{u}ller-Groeling, K. Holinde and J. Speth, Nucl. Phys. {\bf  A  513} (1990) 557.
\bibitem{Weise:96} T.~Waas, N.~Kaiser, and W.~Weise,  Phys. Lett. {\bf B 365} (1996) 12; Phys. Lett. {\bf B 379} (1996) 34; N.~Kaiser, P.~B.~Siegel, and W.~Weise, Nucl. Phys. {\bf A594} (1995) 325; W.~Weise, Nucl. Phys. {\bf A610} (1996) 35.
\bibitem{Shevchenko:06} N.V. Shevchenko, A. Gal and J. Mares, Phys. Rev. Lett. {\bf 98} (2007) 082301.
\bibitem{Ikeda:06} Y. Ikeda and T. Sato, arXiv:nucl-th/0701001.
\bibitem{Akaishi:86} Y. Akaishi, Int. Rev. Nucl. Phys. {\bf 4} (1986) 259.
\bibitem{Tamagaki} R. Tamagaki, Prog. Theor. Phys. {\bf 44} (1970) 905.
\bibitem{Bethe:71} H.A. Bethe, Ann. Rev. Nucl. Sci. {\bf 21} (1971) 93.
\bibitem{DISTO} M. Maggiora {\it et al.}, Nucl. Phys. {\bf A691} (2001) 329c.
\bibitem{FOPI} GSI-FOPI group, FOPI proposal at GSI (2006-2007).

\bibitem{Nambu}Y.~Nambu and G.~Jona-Lasinio, Phys. Rev. {\bf 122}
(1961) , 345 {\bf 124} (1961)  246.
\bibitem{Hatsuda:85} T. Hatsuda and T. Kunihiro, Phys. Rev. Lett. {\bf 55}
(1985) 158-161; Prog. Theor. Phys. {\bf 74} (1985) 765; Prog. Theor. Phys.
Suppl. {\bf 91} (1987) 284-298; Phys. Reports {\bf 247} (1994)  221.
\bibitem{Vogl:91} U. Vogl and W. Weise, Prog. Part.
Nucl. Phys. {\bf 27} (1991)  195.
\bibitem{Kaplan:86} D.B.~Kaplan and A.E.~Nelson, 
Phys. Lett. {\bf B175} (1986) 57. 
\bibitem{Brown:94} G.E. Brown, C.H. Lee, M. Rho and V. Thorsson, Nucl. Phys. {\bf A567} (1994) 937; G.E. Brown, {\it ibid.} {\bf A574} (1994) 217c; G.E. Brown and M. Rho, Phys. Rep. {\bf 269} (1996) 333.


\end{thebibliography}
\end{document}